\documentclass[10pt,aps,prl,twocolumn,notitlepage,floatfix,
               nofootinbib,superscriptaddress]{revtex4-1}

\usepackage{amsmath, amsthm, amssymb, amsfonts, amsbsy, mathrsfs}
\usepackage{bm}
\usepackage{stmaryrd}

\usepackage[hidelinks,bookmarks=true]{hyperref} 
\hypersetup{pdfstartview=FitH,pdfhighlight=/O,colorlinks=false}

\begin{document}

\title{ Gravity limits the kinetic energy of a massive elementary 
        particle }

\author{Justin C. Feng}
\email{jcfeng@ntu.edu.tw}
\affiliation{Leung Center for Cosmology and Particle Astrophysics,
National Taiwan University, No.1 Sec.4, Roosevelt Rd., Taipei 10617,
Taiwan, Republic of China}

%=======================================================================
%-----------------------------------------------------------------------
%
%       ABSTRACT
%
%-----------------------------------------------------------------------
%=======================================================================
\begin{abstract}
    In this note, I argue that tidal effects generically limit the
    kinetic energy of a single massive elementary particle in the
    vicinity of a compact object. As the kinetic energy is increased,
    the differences in the tidal potential over a Compton wavelength
    will at some point exceed the rest mass of the particle. Above the
    threshold, one expects tidal effects to disrupt single-particle
    states, and in turn, one might expect an incident particle
    scattering off a compact object with an energy significantly
    exceeding the threshold to result in a shower of lower energy
    particles. A calculation reveals that the threshold for neutrinos
    scattering off a $10 M_\odot$ black hole within three Schwarzschild
    radii is roughly $1~\text{GeV}$.
\end{abstract}

\maketitle

%=======================================================================
%-----------------------------------------------------------------------
%
%       MAIN
%
%-----------------------------------------------------------------------
%=======================================================================

Various authors have pointed out that gravitational effects can be
greatly amplified in the frame of an observer moving close to the speed
of light \cite{Pirani:1959,Aichelburg:1970,*Aichelburg:1971,Curtis:1978,
dEath:1978,*dEath:1992,Mashhoon:1993,Plyatsko:2016bee}. In particular,
Mashhoon pointed out that gravitational tidal forces can tear apart
sufficiently fast-moving composite particles \cite{Mashhoon:1992}. One
might wonder what happens for individual \textit{elementary} particles.
In this note, I show that if a single massive elementary particle on a
scattering trajectory exceeds a threshold kinetic energy in the vicinity
of a gravitating compact object, the tidal energy differences across a
Compton wavelength in the frame of the particle exceed the rest energy
of the particle. This suggests that single-particle states are strongly
disrupted above this threshold, and one might expect such a disruption
to result in a multi-particle state at late times.

Consider the scattering of a massive particle from a spherically
symmetric compact object of mass $M$ described by the Schwarzschild line
element (setting $c=G=1$ for now):
\begin{equation} \label{Eq:Schwarzschild}
    ds^2 =-fdt^2+dr^2/f+r^2(d\theta^2+\sin^2\theta d\phi^2),
\end{equation}
with $f=1-2M/r$. The trajectory of the scattered particle is a timelike
geodesic $x^\mu=x^\mu(\tau)$ in the equatorial plane $\theta=\pi/2$,
parameterized by proper time $\tau$. The four-velocity is (the
overdot denotes $d/d\tau$):
\begin{equation} \label{Eq:KillingInv}
    u^a:=\tfrac{dx^a(\tau)}{d\tau}
    =\left(e/f,\dot{r},0,l/r^2\right),
\end{equation}
and the invariants of the motion are:
\begin{equation} \label{Eq:KillingInv}
    \begin{aligned}
    e = \dot{t} f,\qquad 
    l = r^2 \dot{\phi}.
    \end{aligned}
\end{equation}
The conserved quantities $e$ and $l$ are the respective specific (per
unit rest mass) kinetic energy and angular momentum of the particle,
which approach their special relativistic values when $r\gg2M$. The
trajectory $r(\phi)$ in the equatorial plane can be found by performing
an integral constructed from the unit norm condition $u_\mu u^\mu=-1$.
However, for simplicity, I consider only the value of $u^\mu$ at the
point of closest approach, which satisfies the condition $\dot{r}=0$,
and is the point along the scattering trajectory where gravitational
effects are the strongest. It is straightforward to show that the radius
of closest approach $r_0$ is greater than the photon radius ($r_0>3 M$),
and is related to the invariants by the formula:
\begin{equation} \label{Eq:PointofClosestApproach}
    l \sqrt{r_0-2 M} = r_0 \sqrt{(e^2-1) r_0+2 M}.
\end{equation}

Tidal accelerations in the frame of the observer traveling along the
aforementioned geodesic can be computed with the geodesic deviation
equation \cite{Wald,*Pirani:1956tn}:
\begin{equation} \label{Eq:GeodesicDeviation}
    a^\mu
    =
    R{^\mu}_{\rho\sigma\nu}u^\rho u^\sigma z^\nu,
\end{equation}
where $R{^\mu}_{\rho\sigma\nu}$ is the Riemann tensor and $z^\mu$ is a
vector field (satisfying $u_\mu z^\mu=0$) representing the displacement
to a neighboring geodesic that is initially parallel to $x^\mu(\tau)$.
The quantity $a^\mu:= u^\rho \nabla_\rho(u^\sigma \nabla_\sigma z^\mu)$
is the relative acceleration between the two geodesics---here, $u^\mu$
represents a four-velocity field of initially parallel geodesics in the
neighborhood of $x^\mu(\tau)$. Now writing
$z^\mu=\varepsilon\hat{z}^\mu$, where $\hat{z}^\mu$ is a unit vector and
$\varepsilon$ is the separation distance, one can compute the magnitude
of the acceleration in the $\hat{z}^\mu$ direction. For large $e \gg 1$,
and at $r_0$, the transverse acceleration dominates, with (orthonormal
basis) components in the geodesic rest frame \cite{FengReiss:2024}:
\begin{equation} \label{Eq:GeodesicDeviationMag}
    \begin{aligned}
    a_r
    \approx \frac{3e^2 \varepsilon_r}{(s-2) s^2 M^2},\qquad
    a_\theta 
    \approx -\frac{3e^2 \varepsilon_\theta}{(s-2) s^2 M^2},
    \end{aligned}
\end{equation}
where $s:=r_0/M$, $a_r$ and $a_\theta$ are the acceleration magnitudes,
and $\varepsilon_r$ and $\varepsilon_\theta$ are the components of
$z^\mu$ in the respective $r$ and $\theta$ directions. The signs
indicate that neighboring geodesics focus in the $\theta$ direction and
defocus in the $r$ direction. Though the result in Eq.
\eqref{Eq:GeodesicDeviationMag} is valid for any scattering trajectory
in the limit $e \gg 1$, the tidal acceleration can become independent of
$e$ in special instances---a timelike geodesic with a strictly radial
component is one example \cite{Mashhoon:1992,Mashhoon:1987}.

From the tidal acceleration, one can infer a tidal potential by
integrating Eq. \eqref{Eq:GeodesicDeviationMag}; a more general approach
for constructing a tidal potential may be found in \cite{Mashhoon:1975}.
Reintroducing factors of $c$ and $G$, the potential is:
\begin{equation} \label{Eq:Potential}
    \begin{aligned}
    \Phi = - 
    \frac{c^2}{G^2} \frac{3 e^2 (\varepsilon_r^2-\varepsilon_\theta^2)}
    {2(s-2) s^2 M^2}.
    \end{aligned}
\end{equation}
The potential $\Phi$ may be interpreted as a specific potential energy.
Now consider what happens when $\varepsilon_i$ is the characteristic
size of a particle of mass $m$. If the particle is composite, then when
$m|\Phi|$ exceeds the binding energy of the particle, the tidal forces
will break it apart, as pointed out in \cite{Mashhoon:1992}. If the
particle is elementary, then $\varepsilon_i$ is on the order of the
Compton wavelength of the particle, and when $|\Phi| \gtrsim c^2$, the
energy differences over the Compton wavelength $2\pi\hbar/mc$ exceed the
rest energy of the particle. Setting $\varepsilon_\theta=0$, one may
obtain a rough limit for the specific energy:
\begin{equation} \label{Eq:ElementaryLimit}
    \begin{aligned}
    e \lesssim  c^2 s \sqrt{\tfrac{2}{3}(s-2)} 
                \frac{M m}{2\pi{m}_{\text{Pl}}^2},
    \end{aligned}
\end{equation}
where ${m}_{\text{Pl}}=\sqrt{\hbar c/G}$ is the Planck mass.
Fundamentally, particles in flat spacetime are described as
energy-quantized excitations of a quantum field (for a specified
momentum), and one might intuitively expect energy differences exceeding
the rest mass to result in particle creation. However, since the
spacetime geometry is held fixed and admits a timelike Killing vector,
one expects the energy expectation values to be conserved (note also
that $\Phi$ changes sign). Still, these considerations indicate that
beyond this threshold, incident single-particle states are strongly
disrupted by tidal effects, and one might expect such states to
generically evolve into multiparticle states at late times. Energy
conservation then dictates that the resulting particles each have
reduced kinetic energies compared to the incident particle.

Neutrinos scattering near the photon radius of stellar-mass black holes
experience a particularly strong disruption. Neutrinos have a mass sum
on the order of $\sim 0.1~\text{eV}/c^2$
\cite{ParticleDataGroup:2022pth}, and near the photon radius $s=3$ of a
$10 M_\odot$ black hole, Eq. \eqref{Eq:ElementaryLimit} yields (after
multiplying by the mass sum) a threshold for individual neutrinos of
roughly $E \lesssim 0.3~\text{GeV}$, and for $s=6$, this increases to
just over $1~\text{GeV}$. For neutrinos passing within a radius $s=6$ of
a $10^6 M_\odot$ black hole (on the order of Sgr A$^*$), the threshold
increases to $E \lesssim 100~\text{TeV}$ (scaling linearly in $M$), an
order of magnitude lower than the highest energy neutrinos detected by
the IceCube observatory \cite{IceCube:2013cdw,*Padovani:2018acg}.

I comment on some implications for future neutrino observations. One may
expect the scattering of a cosmic ray neutrino passing close to the
photon radius of a black hole and greatly exceeding the threshold energy
to result in a shower of neutrinos with energies lower than the incident
particle, so that the spectrum of neutrinos from the vicinity of a black
hole differs significantly from the background above the threshold. More
generally, one might expect strong tidal effects on neutrinos to result
in some spectral feature near the threshold, whether it be a peak, a
strong suppression above the threshold, or something else. Since the
threshold depends on the mass of individual neutrinos, such spectral
features may permit separate constraints on neutrino masses for each
generation.

I leave the reader with some questions for future investigation. The
first concerns a more rigorous analysis of the scattering process in the
framework of quantum field theory in curved spacetime, but such a task
may be nontrivial due to the highly curved and dynamical spacetime
environment in the Fermi normal frame of the scattering trajectories.
The second concerns the question of an analogous effect for massless
elementary particles (the related issue of wave-particle duality is
discussed in \cite{Mashhoon:1987}). The third concerns the case of
extreme mass ratio scattering between black holes; one might expect no
limit on the kinetic energy or relative velocity for classical black
holes since they can neither break apart nor decay, but the behavior of
a black hole under extreme tidal deformation may require a numerical
analysis. The fourth is whether these considerations yield insights into
the idea that gravitational effects may serve as a universal regulator
for quantum field theory (see \cite{Casadio:2009eh} and Sec. 5.3 of
\cite{Carlip:2022pyh} for brief reviews of such proposals).

%=======================================================================

{\em Acknowledgements --- I thank Jiwoo Nam for discussions on neutrino
physics and acknowledge support from the R.O.C. (Taiwan) National
Science and Technology Council (NSTC) grant No. 112-2811-M-002-132. }

%=======================================================================
%       BIBLIOGRAPHY
%=======================================================================

\bibliography{ref}

\end{document}